\documentclass[11pt]{article}
\usepackage[square]{natbib}
\usepackage{amsmath, theorem, amssymb}

\title{Critique of Feinstein's Proof that $P \neq NP$}
\author{Kyle Sabo \and Ryan Schmitt \and Michael Silverman }
\date{June 14, 2007}

\newtheorem{theorem}{Theorem}[section]

\newtheorem{definition}[theorem]{Definition}

\newcommand{\qed}{\nobreak \ifvmode \relax \else
      \ifdim\lastskip<1.5em \hskip-\lastskip
      \hskip1.5em plus0em minus0.5em \fi \nobreak
      \vrule height0.75em width0.5em depth0.25em\fi}

\newcommand{\records}{\textsc{Find-Record}}
\newcommand{\mim}{\textsc{Meet-In-The-Middle}}
\newcommand{\ssum}{\textsc{Subset-Sum}}
\newcommand{\fthree}{\textsc{First-Three}}

\begin{document}
\maketitle

\begin{abstract}
We examine a proof by Craig Alan Feinstein that $P \neq NP$. We present counterexamples to claims made in his paper and expose a flaw in the methodology he uses to make his assertions. The fault in his argument is the incorrect use of reduction. Feinstein makes incorrect assumptions about the complexity of a problem based on the fact that there is a more complex problem that can be used to solve it. His paper introduces the terminology ``imaginary processor'' to describe how it is possible to beat the brute force reduction he offers to solve the \ssum\ problem. The claims made in the paper would not be validly established even were imaginary processors to exist.
\end{abstract}

\section{Introduction}
In this paper we analyze the argument set forth by Craig Alan Feinstein in his paper, `A New and Elegant Argument that $P \neq NP$' \cite{feinstein2007pnp}.
We present his argument and a counterargument by using his reasoning to ``prove'' a clearly trivial problem to be non-polynomially difficult.

\section{Feinstein's Argument}
Feinstein argues that $P \neq NP$ through the use of ``imaginary processors.'' The author begins by introducing the problem of searching for a record out of $n$ unsorted records. We state this as the following formal language.
\begin{definition}
\records$\ = \{ \langle R, r, i\rangle\ |\ r = r_i \in \{r_1,r_2,...,r_n\} = R\}$
\end{definition}

Feinstein then claims that \records\ is $\Theta(n)$. However, he continues by arguing that using multiple processors to search subsets of $R$ it is possible to achieve a better runtime. The author's definition of ``multiple processors'' is one such that if there were a \records\ algorithm using $n$ processors, it would be possible to find the record in $\Theta(1)$ time.

In this way, Feinstein's use of multiple processors is intuitively similar to a non-deterministic Turing Machine. Each branch of the computation would use another processor. However, the author never makes this comparison and it is possible, therefore, that his model of computation has some extra properties that a traditional nondeterministic Turing Machine does not. We use Feinstein's notation of multiple processors in order to not bias the reader to disagree with his argument during its presentation. We will see that under every possible reasonable definition of multiple processor computation Feinstein's arguments fail to prove his claim.

Feinstein notes that using $n$ processors to solve \records\ is expensive, and that the best solution, he argues, is to instead optimize the sum of processors used and computations per processor. This, under his notation, implies that the best solution is to use $\Theta(\sqrt{n})$ processors, each doing $\Theta(\sqrt{n})$ computations. Furthermore, he implies that the only way to beat the efficiency of brute force is to use these multiple processors.

Feinstein then describes the \ssum\ problem. In  his paper, the \ssum\ problem is the task of deciding if for a given set of integers, $S\ (||S|| = n)$, and another integer, $x$, it is possible to find set $S'\ (||S'|| = m \leq n)\subseteq S$ such that the sum of all the elements in $S'$ is equal to $x$. We can state this alternatively as:

\begin{definition}
\begin{align*}
    \ssum &= \\
&\left\{\langle S, x, y\rangle\ |\ y =
    \begin{cases}
        1 & \textbf{ if there is a set } S' \subseteq \displaystyle S\ |\ x = \sum_{w\in S'} w\\
        0 & \textbf{ otherwise }
    \end{cases}
    \right\}
\end{align*}
\end{definition}

To solve the \ssum\ problem, one can simply check all possible subsets, which would take $\Theta(2^n)$ time, however, Feinstein notes there are faster solutions, such as the \mim\ algorithm which is $O(\sqrt{2^n})$. Feinstein argues that it is possible to beat the brute force approach of \records\ because there is an inherent mathematical structure to \ssum. That is, there is some property the \ssum\ problem has that the \records\ problem does not have.

Essentially, Feinstein argues that \ssum\ is a search for a record where instead of a special number or label, the record is a subset with the special property of having a sum equal to the desired value $x$. This special record is a specific subset or set of subsets of $S$ out of all possible subsets. There are clearly $2^n$ subsets, where again $n$ is the size of $S$, and therefore the \records\ search through each subset will take $\Theta(2^n)$. More formally, the paper offers a non-polynomial time reduction to \records\ from \ssum, though the paper never formally defines the reduction. Given an instance of \ssum\, Feinstein describes how to enumerate all subsets and their sums, $\Theta(2^n)$, and then searches that list for a record that has a sum of $x$. However, as Feinstein notes, it is clear the algorithm \mim\ has a runtime of $O(\sqrt{2^n})$.

To reconcile this, Feinstein states there must be some inherent mathematical structure to \ssum\ which allows for a faster solution than using the reduction to \records. Using the reduction above, in essence Feinstein implies that the since the \records\ search is  $\Theta(e)$, where $e$ is the number of elements to search through, and that since $S$ has $2^n$ subsets, where $n$ is the size of $S$, it is impossible to beat the runtime $O(2^n)$ without somehow searching the $2^n$ subsets of $S$ in faster than linear time.

Since there is in fact a way to solve \ssum\ faster than such a linear search through the $2^n$ subsets of $S$, and since, according to Feinstein, the only way to go faster is to search through the $2^n$ subsets in less than $2^n$ operations, there must be some aspect of the \mim\ algorithm that recreates multiple processors which Feinstein calls imaginary processors.  That is, since checking every possible subset would take longer than \mim\, there must be some way that \mim\ achieves this speedup. Since there is an inherent mathematical structure to \ssum, according to Feinstein, \mim\ can create these ``imaginary processors'' that give the same boost in computation that using extra real processors does for \records.

Feinstein then states that there is a computational penalty to creating these $\Theta(\sqrt r)$ imaginary processors, where $r$ is the number of records to search through and we have each of the $\sqrt r$ imaginary processors searching $\sqrt r$ elements. He argues that because each creation requires a sorting operation, the amount of time it takes to create each ``imaginary processor,'' is $\Theta(p)$ where $p$ is the number of processors. Therefore, he argues, we are bound again by minimizing the product of the number of `imaginary processors,' times the number of elements processed by each, giving us a lower bound of $\Theta(\sqrt{2^n})$. Feinstein states that since \ssum\ is now bounded to $\Theta(\sqrt{2^n})$, the best solution for an $NP$ complete problem requires more than polynomially many computations and therefore $P \neq NP$. This bound is based entirely on Feinstein's idea that it is too expensive to create more imaginary processors and that these imaginary processors are the only way to get a speedup.

\section{Brief Summary}
In the earlier section we present Feinstein's argument in as rigorous and formal a way as possible without changing the spirit or meaning of his argument. We now reword his argument to the equivalent and more simple logically valid statements.

\begin{enumerate}
\item A brute force search for \ssum\ takes $\Theta(2^n)$ unless we use multiple processors. This is because we have a non-polynomial time reduction from \ssum\ to \records\ and records cannot be solved faster than linear time without multiple processors.
\item \mim\ is $\Theta(\sqrt{2^n})$
\item \mim\ must simulate multiple processors by creating imaginary processors to achieve better than $\Theta(2^n)$ because of 1 and 2.
\item The only way to get a faster runtime for \ssum\ without multiple processors is to use imaginary processors.
\item The best runtime possible for \ssum\ with imaginary processors is $\Theta(\sqrt{2^n})$  due to penalties of creating new imaginary processors.
\item Since 4 and 5, $P \neq NP$.
\end{enumerate}

\section{Counterargument}
The fault in Feinstein's paper lies in neglecting proof of some nontrivial statements that are implied by his argument. Most simply, the paper fails to prove that the reduction given in argument 1 above is the most efficient reduction. The author argues that since the reduction shown to exist in argument 1 above is $\Theta(2^n)$, any such algorithm that has faster than $\Theta(2^n)$ runtime for \ssum\ must be imitating some more powerful form of computation. This is simply not true. There could in concept be a more efficient reduction that could even be polynomial time. Without proof that the reduction in $3$ is the most efficient such reduction, it is impossible to make the author's claim.

Even so, let's assume the reduction in $3$ is optimal. The author's argument then relies on the fact that the non-polynomial solution of brute force has a non-polynomial runtime to imply that the problem the non-polynomial time algorithm being used to solve must somehow be using imaginary processors.

If Feinstein were correct, then many things use imaginary processors. For instance, consider checking if the first number in a list of $n$ numbers is 3, where $n$ varies on input, called \fthree. We offer a non-polynomial reduction from \fthree\ to records. We can take the first number of our input list, and put it in a list of $n$ random but non-3 elements, and then we can repeat putting the element into new random lists $n!$ times. This algorithm is a non-polynomial reduction from \fthree\ to \records. Therefore, according to Feinstein's reasoning, \fthree\ not only cannot be done in polynomial time, but to do better than the speed of the reduction it must use imaginary processors. However, it seems clear that the code that outputs true if the first element is 3, does not use imaginary processors, and is constant time. If Feinstein were correct, \fthree\ could not be solved this way.

\bibliographystyle{alpha}
\bibliography{SaboSchmittSilverman}
\appendix
\end{document}